# LOW THRESHOLD PARTICLE ARRAYS

KONRAD BERNLÖHR

*Max-Planck-Institut für Kernphysik*
*P. O. Box 103980, D-69029 Heidelberg, Germany*

**Abstract.** While atmospheric Cherenkov telescopes have a small field of view and a small duty fraction, arrays of particle detectors on ground have a 1 sr field of view and a 100% duty fraction. On the other hand, particle detector arrays have a much higher energy threshold and an inferior hadron rejection as compared to Cherenkov telescopes. Low threshold particle detector arrays would have potential advantages over Cherenkov telescopes in the search for episodic or unexpected sources of gamma rays in the multi-TeV energy range. Ways to improve the threshold and hadron rejection of arrays are shown, based on existing technology for the timing method (with scintillator or water Cherenkov counters) and the tracking method (with tracking detectors). The performance that could be achieved is shown by examples for both methods. At mountain altitude (about 4000 m or above) an energy threshold close to 1 TeV could be achieved. For any significant reduction of the hadronic background by selecting muon-poor showers a muon detection area of at least 1000 m$^2$ is required, even for a compact array.

**Key words:** cosmic rays – gamma rays – air showers

## 1. Introduction

The use of arrays of particle detectors for observing cosmic rays of PeV energies and above has a long history. Scintillation counters are the traditional type of detectors for these arrays. The arrival directions of the extensive air showers and, thus, the primary cosmic rays are derived from timing measurements between the counters. Reports of an excess of cosmic rays ($\gamma$-rays) from the direction of Cygnus X-3, modulated with the 4.8 h orbital period of the binary system (Samorski and Stamm, 1983; Lloyd-Evans et al., 1983), and subsequent reports of sporadic emission from several candidate sources stimulated the build-up of several new arrays of much improved sensitivity and lower energy threshold (typically 30–100 TeV). Despite the improved sensitivity no clear $\gamma$-ray source detections have been achieved with these new arrays.

At lower energies, around 1 TeV, the imaging air Cherenkov technique with a significant reduction of the almost isotropic background of protons and nuclei has been applied with success. One source, the Crab nebula, has been observed by several experiments and a few others have been detected by single experiments at quite high significance level (e.g. see Weekes, 1992, and references therein; Baillon et al., 1993; Goret et al., 1993).

In comparison with air Cherenkov telescopes the particle detector arrays have conceptually several important advantages. The particle arrays can be operated at day and night while the air Cherenkov systems (both telescopes and wide-angle counter arrays) are restricted to clear, moonless nights, which is about 10–15% of the total time at excellent sites. For Cherenkov systems any candidate source is too close to the sun to be observable at night for about four to six months each year. Arrays observe on- and off-source regions at the same time while telescopes, at least so far, are splitting their valuable observing time between on- and off-source. Arrays can even observe several candidate sources at the same time, due to their large field of view, about 1 sterad. The Cherenkov telescopes have a 100-300 times smaller field of view.

On the other hand, existing particle arrays have higher energy thresholds than Cherenkov telescopes (10–1000 TeV compared to 0.3–3 TeV). The rejection of the hadronic background by particle arrays is much inferior to that achieved by image analysis of imaging air Cherenkov telescopes, at least near threshold energies. The hadron rejection of arrays can improve substantially with increasing energy while that of Cherenkov telescopes hardly improves or even deteriorates. Due to the attenuation of extensive air showers, low threshold particle arrays should be located at rather high mountain altitude. The requirement of large flat areas is, at mountain altitude, another potential disadvantage of particle arrays.

Future particle arrays with thresholds of a few TeV, which could be called low threshold particle arrays in the context of this workshop, would for the reasons mentioned – mainly the inferior hadron rejection – not be built to search for gamma-rays from a few candidate point sources. Atmospheric Cherenkov telescopes are expected to remain unrivalled in this respect. Low threshold particle arrays could be of better use to search for episodic sources with 'bursts' on all time scales (seconds to months), to search for unexpected sources, and to examine source spectra in the multi-TeV range.

Although the planned MILAGRO experiment will be a low threshold particle array, it is not covered by this paper but is addressed in detail by another paper (G. B. Yodh, these proceedings).

## 2. Existing arrays and requirements for source detections

Before presenting the existing arrays it is instructive to evaluate actual requirements for observing known VHE gamma-ray sources with a particle array. In a simplified way, assuming Gaussian statistics and neglecting the problem of off-source background determination, the signal-to-noise ratio of an array for a point source is proportional to

$$S/N \propto \frac{1}{\sigma}\sqrt{\frac{At}{r_{\rm h}}},$$

with the angular resolution $\sigma$, the effective area $A$, the measuring time $t$, and the hadron rejection $r_{\rm h}$. The hadron rejection depends on the fractions $f_\gamma$ of gammas and $f_{\rm h}$ of hadrons passing the trigger and experimental shower selection: $r_{\rm h} = f_\gamma^2/f_{\rm h}$. In general, $\sigma$, $A$, and $r_{\rm h}$ are functions of energy.

A detailed calculation, assuming a resolution of $\sigma = 0.5°$, a circular source search bin of $1.6\,\sigma$ radius and taking into account the measured fluxes of the cosmic ray background and the Crab nebula results in the following requirements for an array that should be able to observe the Crab at a 5 sigmas significance level in one year with 10000 s observing time each day: Without any hadron rejection an effective area of at least $6 \cdot 10^4$ m$^2$ above 1 TeV or $7 \cdot 10^5$ m$^2$ above 10 TeV would be required. In alternative terms the required hadron rejection for an array of $10^4$ m$^2$ effective area would be at least 6 above 1 TeV or at least 80 above 10 TeV.

Existing scintillator arrays fall substantially behind these requirements. Table I shows a selection of scintillator arrays with fairly low threshold energies $E_{\text{thresh}}$ and good angular resolution. The quoted thresholds should be used with care because different groups use different threshold definitions. Where available, the thresholds for gamma-rays have been used. Convenient definitions are the mode energy of detected gammas (or protons) or the energy where the array reaches 50% detection efficiency for showers with cores inside the array boundaries. Where possible, the later definition is applied throughout this paper.

TABLE I

Air shower scintillator arrays with $E_{\text{thresh}} \leq 100$ TeV and angular resolution of 1° or better.

| Array | Threshold (TeV) | Depth (g cm$^{-2}$) | No. of stations × area (m$^2$) | Muon det. area (m$^2$) | Array area ($10^4$ m$^2$) |
|---|---|---|---|---|---|
| CASA-MIA[1] | 70 | 870 | 1089 × 1.5 | 2550 | 23 |
| CYGNUS-I[2] | 50 | 800 | 108 × 0.8 | 110 | 2.2 |
| EAS-TOP[3] | 100 | 800 | 29 × 10 | — | 10 |
| HEGRA[4] | 40 | 800 | 257 × 1 | 270 | 4 |
| JANZOS[5] | 100 | 850 | 45 × 0.5 + 31 × 1 | — | 2 |
| SPASE[6] | 50 | 700 | 24 × 1 | — | 0.6 |
| Tibet AS$\gamma$[7] | 10 | 600 | 49 × 0.5 + 16 × 0.25 | — | 0.8 |
| ( Tibet II[8] | ≤10 | 600 | 221 × 0.5 | — | 3.7 ) |

[1] Borione et al., 1994 – [2] Alexandreas et al., 1991 – [3] Aglietta et al., 1989
[4] Aharonian et al., 1993 – [5] Allen et al., 1993 – [7] Beaman et al., 1993
[7] Amenomori et al., 1993b – [8] Yuda, 1994

Of the arrays shown in tab. I, only the Tibet AS$\gamma$ array (Amenomori et al., 1993b) can be called a low threshold array in the context of this conference. It has a threshold energy of about 10 TeV and is currently being upgraded into the Tibet II array with a substantially larger area but little reduction in threshold (see fig. 1, Yuda, 1994). The Tibet II array should be operated with a 200 Hz trigger rate when it is complete in 1995. Neither the old nor the new Tibet array have any hadron rejection other than that due to different threshold energies for gammas and cosmic ray nuclei. Arrays with the potential for a substantial reduction of the hadronic background, in particular CASA-MIA (Borione et al., 1994) by selecting muon-poor showers and also HEGRA (Aharonian et al., 1993) by using muons and by other methods, have a much larger threshold energy than the Tibet array.

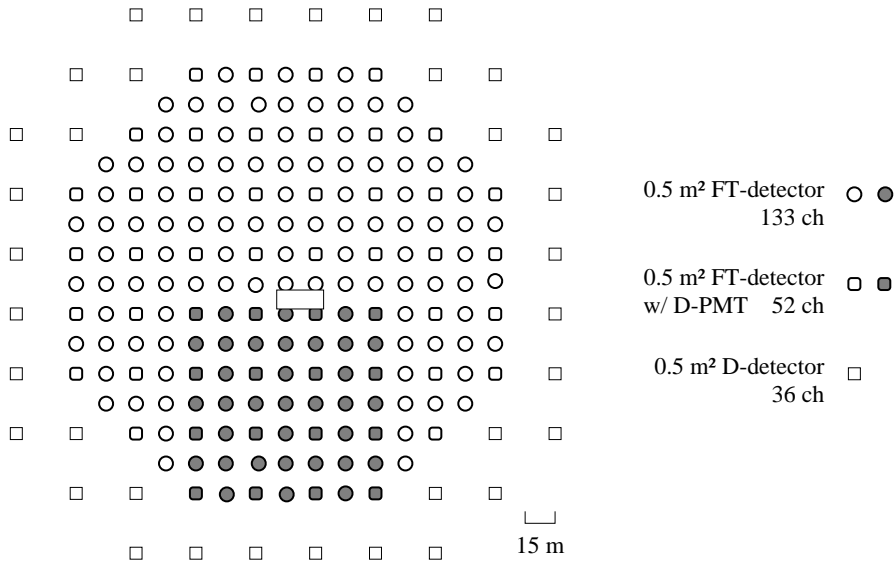

Fig. 1. The new Tibet array (figure kindly provided by T. Yuda). Shaded symbols indicate scintillation counters from the old Tibet AS$\gamma$ array. The active array area with 185 fast-timing counters will be surrounded by additional 36 density counters.

## 3. Hadron rejection with particle arrays

The most powerful method for rejecting hadronic showers with particle arrays is to distinguish the showers by their muon content. The average muon content of a gamma shower is expected to be about a factor of 30 less than that of a proton shower of the same energy. The muon content of showers initiated by heavy nuclei is even larger than that of proton showers. At energies of a few TeV just a few tens to a few hundreds of muons are expected to arrive on ground. The detection of a single shower muon should, therefore, be sufficient to reject a hadronic shower.

The detection of the muons from air showers is usually done with particle detectors placed under a thick absorber layer of dirt, lead, or another material in which essentially all particles other than muons, in particular $e^+/e^-$, should be absorbed. Several practical problems arise when dealing with the detection of just a few muons by low threshold arrays: a) random muons, b) detector noise, c) natural radioactivity, and d) shower electrons misidentified as muons (*punch-through electrons*). Random muons arrive at a rate of 200–300 m$^{-2}$ s$^{-1}$ on mountain altitude, depending on altitude and geomagnetic latitude. Rates due to photomultiplier noise and radioactivity are frequently well above 1 kHz per PM (6 kHz in the case of the CASA-MIA muon detectors, see Borione et al., 1994).

The best cure for these problems (a–d) would be achieved by measuring the direction of the muons which should be almost parallel to the shower axis for genuine shower muons. For scintillation counters or other direction-insensitive detectors one could use coincidences of several layers to resolve the noise problems (b and c), although at a multiple of the cost. A much more cost-effective way to reduce problems a–c is to accept only those muon counters which fired within a short time interval at the shower front. A thicker absorbers would be useful to reduce punch-through electrons (d) but ultimately also absorbs many muons. The usual compromise is an absorber thickness of about 20–25 radiation lengths, corresponding to a cutoff energy for vertical muons of about 1 GeV.

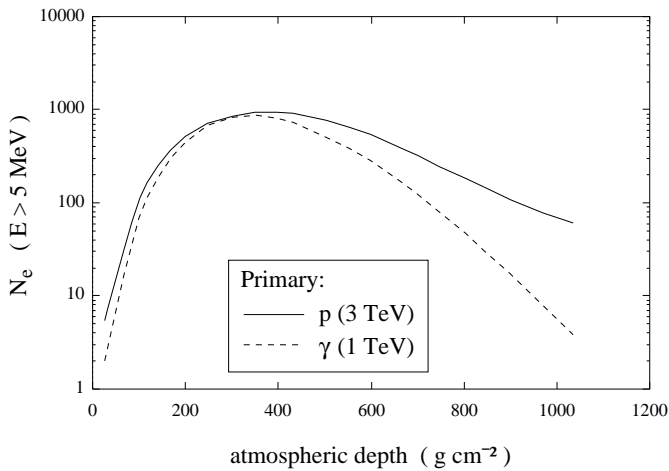

Fig. 2. Average longitudinal development of a 3 TeV proton shower (solid line) and a 1 TeV gamma shower (dashed line) in terms of the number of $e^+/e^-$ with energy above 5 MeV (numerical simulation). Note that, although the 1 TeV gammas have almost the same peak shower size as protons of three times larger energy, the gamma showers are attenuated much faster than the proton showers.

The different longitudinal development of electromagnetic and hadronic showers in the atmosphere (see fig. 2) implies that for arrays situated at high altitudes (above about 4000 m) a substantially lower threshold energy – by more than a factor of two – for gammas than for protons can be achieved. Due to the steep cosmic ray spectrum a reduction of the proton background by about a factor of 3 is achieved. Thresholds for heavier primaries are even higher than for protons. Iron nuclei, for example, would be suppressed by an additional factor of about 6 with respect to protons. At altitudes of 4000–5000 m the composite hadronic background is reduced by a factor of 4–6 with respect to any gamma-ray signal. At an altitude below 2000 m a proton of a few TeV primary energy results, on average, in a larger shower than a gamma of the same energy. A low threshold array at low altitude, as had been proposed for ARGO (Catalano et al., 1992), would thus be enriched with hadronic showers instead of gamma showers.

Other ways to discrimate between hadronic and gamma showers could take advantage of differences in some average shower properties, e.g. the lateral distribution, including fluctuations and asymmetry thereof, or different energy spectra of the secondary electrons. Due to the shower-to-shower fluctuations and the small number of secondaries detected per shower, these other ways are not effective at low threshold. Muon detection and a high-altitude location are the only conceivable ways to reduce the hadronic background with low-threshold particle arrays.

## 4. Ways to reduce the thresholds of particle arrays

Among the main improvements from the arrays of the early 1980s to those of the early 1990s is a ten- to hundred-fold increase in the number of counters per array. Most present large arrays have more than 100 counters, CASA-MIA even uses more than 5000 photomultipliers. Although the 1% area fraction covered by counters in an array seems to leave plenty of room for further improvements, current costs of more than 10 mio. US$ for a large array lead us to consider more cost-effective ways to improve the sensitivity than just increasing the number of counters.

In air showers at ground level there are roughly a factor of five more secondary gammas than $e^+$ and $e^-$. Even near the core there are more than three times as many gammas as electrons. With scintillation counters a lead (or another high $Z$ material) conversion layer above the scintillator causes most energetic gammas to convert to $e^+e^-$ pairs while low-energy electrons are absorbed in the lead. The best compromise between conversion and absorption is a converter thickness of about 1 radiation length. Average signal amplitudes from showers increase by a factor of 1.6. The time resolution and, thus, the angular resolution improves by even a factor of 1.7 because gammas have, on average, a smaller time delay with respect to the shower front than electrons. However, the number of fired stations is little increased and thresholds are only decreased by about 20%. In all arrays of table I this converter technique is already applied.

Water Cherenkov detectors attempt to make use of the signal from converted gammas without loosing that of many electrons. Due to the low light output of the Cherenkov process (at most 100 visible photons per MeV deposited energy, about 1% of the light output from plastic scintillators) and due to the directed Cherenkov emission a very large number of large photomultipliers is needed to detect most of the particles impinging into the water. Without reflectors and multiple scattering an 8-inch PM one meter below the surface would register just about ten photoelectrons from a 10 MeV electron impinging onto a $1\,\mathrm{m}^2$ area ring on the surface. Nevertheless, a considerably lower threshold energy can be achieved by the water Cherenkov technique than is achieved with existing scintillator arrays (G. B. Yodh, these proceedings).

In the case of a large scintillator array one could reduce the threshold by reducing the spacing between stations. A reduction of the spacing by a factor of three, for example from a typical 15 m to a 5 m separation, results in an array where the same number of counters are hit in showers of about three times less energy with, on average, more particles per counter in the smaller scale array. It is estimated that such a reduction in scale would reduce the threshold by a factor of three to, at most, six at 1/9 of the old array area and a 2–3 times worse angular resolution at the reduced threshold. Therefore, the optimum separation is a question of the expected source spectrum. See Gibbs et al. (1988) for the optimization of the CASA array.

As figure 3 illustrates, current state-of-the-art scintillator arrays have thresholds in terms of minimum shower size ($N_\mathrm{e}$) of 5000-10000 and differences in threshold energy are mainly due to different altitudes. A high-altitude location is obviously very important for achieving a low threshold energy. A further increase in altitude beyond that of the Tibet array, however, cannot reduce threshold energies of scintillator arrays much more. Figure 3 also indicates the much lower threshold that could be achieved by an array of tracking detectors, like the proposed Cosmic Ray Tracking (CRT) array (Heintze et al., 1989).

## 5. Arrays of tracking detectors

Arrays of tracking detectors take advantage of the fact that the arrival directions of secondary particles in air showers are well correlated with the direction of the primary. A good angular resolution can, therefore, be achieved with few detected secondaries. An estimate of the possible shower resolution in each projection is

$$\sigma \approx (3° \text{ to } 4°)/\sqrt{n},$$

where $n$ is the number of detected secondaries. No large spacing between the detectors is needed to achieve that resolution, except that the array should be large enough to locate the shower core. Compact arrays of tracking detectors can be built, if a low threshold is desired. Tracking detectors also have the potential for excellent muon identification. A single detected shower muon can be sufficient to reject a hadronic shower.

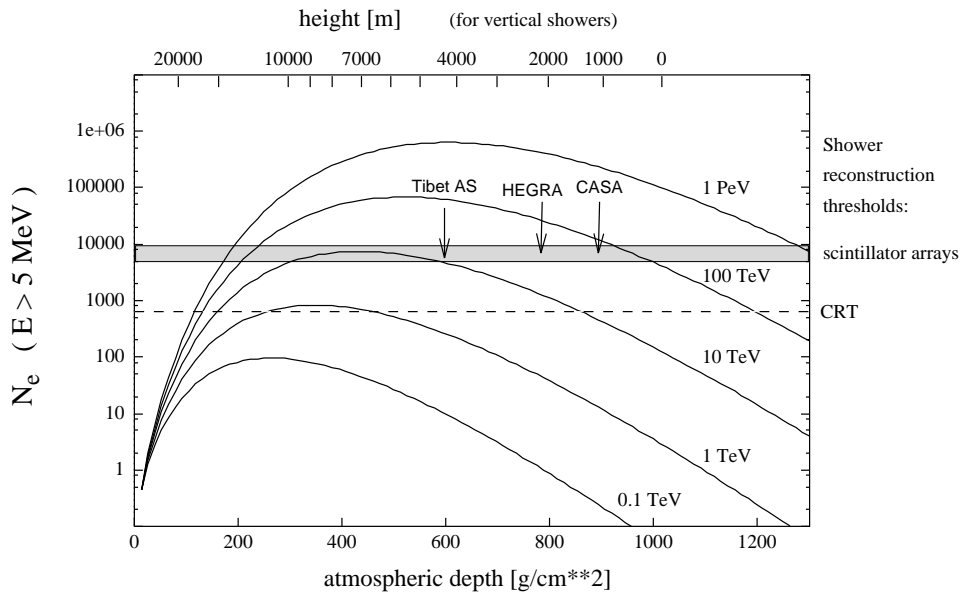

Fig. 3. Average longitudinal development of gamma showers of 0.1 TeV to 1 PeV in terms of the number of $e^+/e^-$ with energy above 5 MeV (analytical approximation). Shower reconstruction thresholds of several scintillator arrays (see table I) in terms of minimum shower size are indicated by the shaded band. Arrays of tracking detectors like CRT could achieve lower thresholds (dashed line).

The first operational array of tracking detectors is GRAND (Poirier et al., 1988, Poirier et al., 1993), consisting of 64 stations equipped with multiwire proportional chambers (MWPCs). Each station has a stack of three pairs of MWPCs of 1.2 m$^2$ area above a 5 cm thick steel plate to reconstruct tracks and one pair of MWPCs below to identify muons. Single-track angular resolutions in each projection of 0.26° for muons and 0.35° for electrons are achieved. A subarray of at least 16 stations has been operational for a while and the whole array should be complete by the end of 1994 (Poirier, 1994). In some sense, GRAND is a real low-threshold array for hadronic cosmic rays, due to a single-muon trigger in addition to an air-shower trigger. Event rates of 1500 Hz are anticipated for this array.

For two reasons, at least, GRAND cannot (and is not intended to) be a low threshold array for gamma-ray astronomy. The first reason is its low altitude (220 m) and the second is its rather poor rejection of punch-through electrons. Due to the thin absorber plate and the single pair of MWPCs below it, more than 3% of the shower electrons are misidentified as muons.

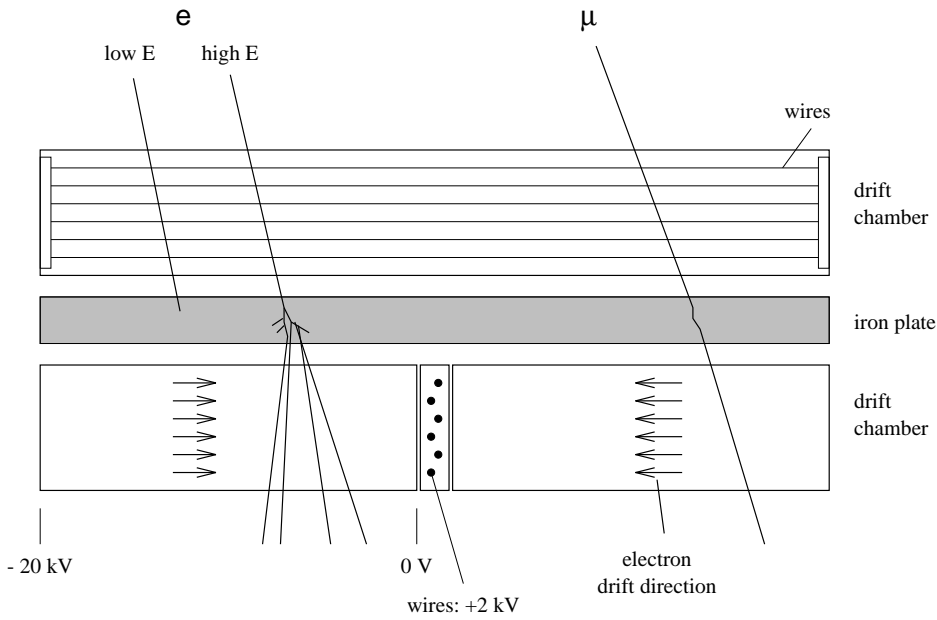

Fig. 4. Schematical view of a CRT detector from the side. Drift chambers and iron plate have a circular cross section of 1.8 m diameter. The wire direction in the upper drift chamber is perpendicular to that in the lower chamber, for easy calibration of the drift velocity. Muons are identified by a matching pair of tracks in each chamber, with little scattering in the iron plate. Electrons are either absorbed in the iron plate or cause an electromagnetic shower extending to the lower chamber.

Another type of tracking detectors, the CRT detectors achieves a much better rejection of punch-through electrons. CRT detectors consist of two circular drift chambers of 1.8 m diameter and a 10 cm thick steel plate inbetween (fig. 4). Each drift chamber has two D-shaped drift cages made of printed-circuit boards (PCBs) and a proportional wire chamber in the middle, with 6 sense wires of more than 1.8 m length supported by an aluminium frame. The drift chambers and the iron plate are in a gas-tight vessel filled with argon-methane. The readout of the wires and segmented-cathode strips is done by a 40 MHz FADC system. Each detector has its own CPU for readout and track reconstruction. For more details and the performance of CRT detectors see Bernlöhr et al. 1993a,b. Because of the thicker absorber plate and a full track reconstruction in the lower chamber, only 0.07% of the shower electrons are misidentified as muons. Therefore, hadronic showers can be identified by single muons.

Ten CRT detectors are in operation at the site of the HEGRA array (Aharonian et al., 1993) on La Palma. By comparison with showers reconstructed from HEGRA array data, a CRT single-muon angular resolution per projection of about 0.4° is determined. Due to the excellent shower muon identification and the good angular resolution the CRT detectors, together with the HEGRA array, seem to be capable to measure the average composition of cosmic rays near the knee of the energy spectrum by the muon angles with respect to the shower axis. The search for gamma-ray sources is beyond the scope of the present small-scale array of ten CRT detectors.

## 6. Comparison of potential low threshold arrays

The capabilities of different detector techniques can be illustrated by examples of potential arrays. The examples also show by how much present particle detector arrays have to be improved to be able to detect a known gamma-ray source like the Crab nebula. The examples are based on full numerical shower simulations with CORSIKA (Capdevielle et al., 1993), including a simple treatment of detection efficiencies of the different types of detectors for different types of particles (electrons, gammas, muons). Triggering of the example arrays has been included in the simulation but no reconstruction of 'detected' showers.

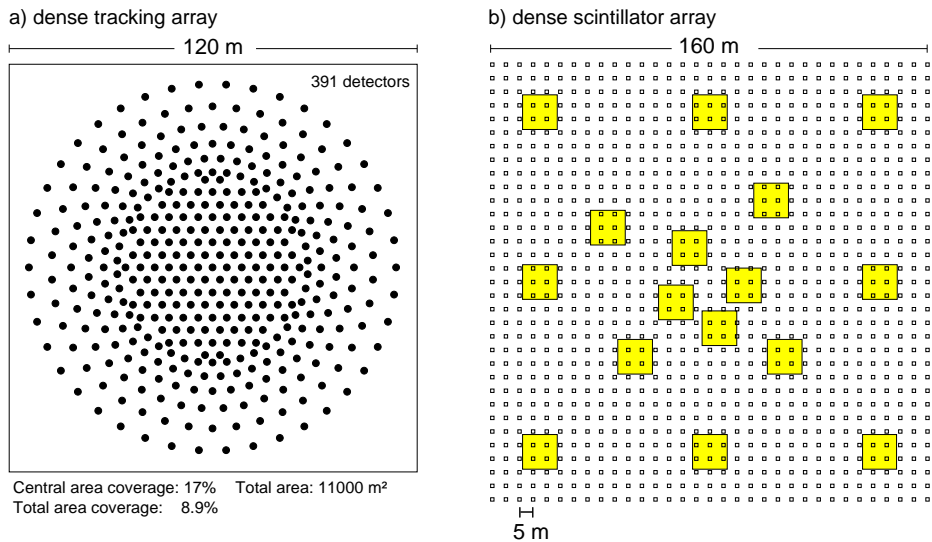

Fig. 5. Potential arrays used in comparison: a) dense CRT array of 391 detectors in a circular area of 11000 m$^2$, b) dense scintillator array of 1089 surface counters of 1.44 m$^2$ each and 16 patches of 64 underground muon counters each (shaded squares) of 160 m$^2$ each.

In order to compare with the existing array with the lowest threshold, the Tibet array, all examples were calculated for the altitude of this array, 4300 m. Arrays of the size and layout of the Tibet array (old and new enlarged version) have been included in the simulations but trigger conditions differ from those used at the real experiment.

Two potential low threshold arrays not actually planned were used for this comparison (see figure 5): a CRT array which has a number of detectors comparable to the original design (Heintze et al., 1989) but is much more compact to achieve a low threshold, and a dense scintillator array. Sizes and numbers of the scintillation counters correspond to the CASA array but separations are only a third of those in CASA. In contrast to CASA, the simulations assume that each station could be fired by a single ionizing particle as, for example, in the Tibet or HEGRA arrays.

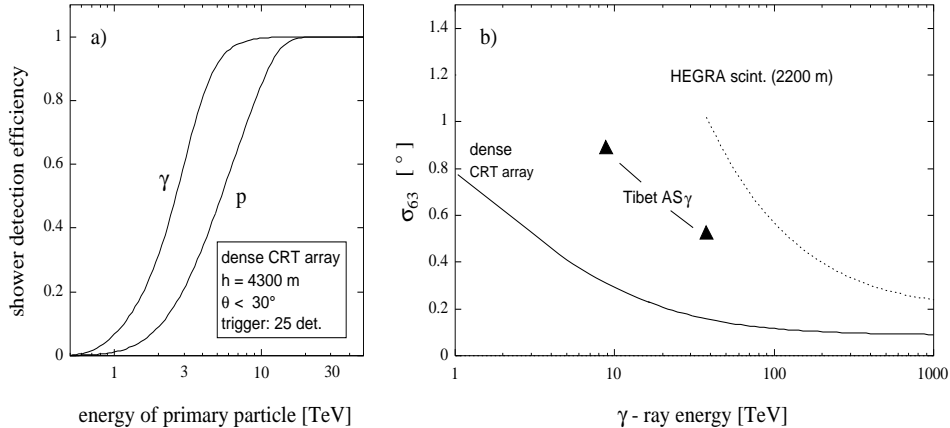

Fig. 6. a) Fraction of detected gamma-ray and proton showers with shower core inside a potential dense CRT array if a 25-of-391 stations trigger is used. b) Estimated resolution of a dense CRT array for $\gamma$-ray showers compared with the existing Tibet (Amenomori et al., 1993a) and HEGRA (Aharonian et al., 1993) scintillator arrays.

Figure 6a shows the efficiency of detecting gamma-ray and proton showers with the CRT array, assuming a 25 stations trigger. The efficiency curves of the scintillator array are quite similar and thresholds are comparable to those of the CRT array. In both cases the threshold for gamma showers is almost a factor of two lower than for proton showers. For gammas the 50% efficiency is reached at about 2 TeV for all showers of less than 30° zenith angle and at less than 1.5 TeV for vertical showers. The angular resolution of the potential CRT array has been estimated for the detected showers (fig. 6b). That of the scintillator array is more difficult to estimate without detailed detector simulations but is most probably worse than that of the CRT array. For underground muon detectors made of scintillation counters it would be difficult to get the noise rate of 1024 counters small enough to reject hadronic showers on the basis of one fired counter without loosing most genuine gamma showers as well.

Within the uncertainties of the estimated angular resolutions and hadron rejections, the sensitivities of the assumed dense tracking and scintillator arrays are comparable. Assuming the Crab flux spectrum of Lewis et al.

TABLE II

Potential low threshold arrays assumed at the altitude of the Tibet array (4300 m)

|  | Tibet II[1] | dense CRT | dense scint. |
| --- | --- | --- | --- |
| No. of timing or tracking detectors | 185 | 391 | 1089 |
| Area of timing/tracking detectors | 0.5 m$^2$ | 2.5 m$^2$ | 1.5 m$^2$ |
| Total muon detection area | — | 980 m$^2$ | 2550 m$^2$ |
| 50% efficiency for $\gamma$ showers | 9 TeV | 2.5 TeV | 2.5 TeV |
| 50% efficiency for proton showers | 17 TeV | 5.3 TeV | 5.0 TeV |
| Total area of array | 36 900 m$^2$ | 11 000m$^2$ | 25600 m$^2$ |
| Hadron rejection by muon-poor showers (in addition to that due to thresholds): | | | |
| All detected events | — | 3 (4)[2] | 1.5/4 (2/9)[2,3] |
| Above 3 TeV | — | 5 (7)[2] | 2/6 (4/18)[2,3] |
| Above 10 TeV | — | 12 (30)[2] | 3/16 (60)[2] |

[1] Simulated trigger and, therefore, thresholds differ from those of real experiment.
[2] Numbers in parenthesis: only showers with cores near array centres.
[3] At least 3 muons / at least 1 muon required to reject hadronic shower.

(1993), a 4–5 sigmas significance could be achieved with both types of arrays in about 1–2 years of measurements. Because the improvement in hadron rejection with increasing energy (or shower size) approximately compensates the decreasing flux, the achieved significance level would be quite independent of any energy (or size) cut in the 2–20 TeV range. A possible short-term enhancement as reported for Cygnus X-3 (Dingus et al., 1988) would be detectable within one day. The Tibet array, even after its present enlargement, is not expected to see a significant signal from the Crab nebula, due to a worse resolution and missing hadron rejection by muon-poor showers.

## 7. Conclusions

Although the thresholds of particle arrays have been lowered by more than an order of magnitude during the last decade, there is still a significant gap between the energy ranges of Cherenkov telescopes and of particle arrays. The existing array with the lowest threshold of about 10 TeV, the Tibet array, is not expected to see a significant signal from the Crab nebula.

With new, denser particle arrays which could be built with tracking detectors or scintillation counters, as demonstrated in the previous section, or with a water Cherenkov system like MILAGRO, thresholds can be reduced close to 1 TeV while having a significant hadron rejection. In order to achieve any hadron rejection at a few TeV, a muon detector area of at least about 1000 m$^2$ is required, even for a compact (10$^4$ m$^2$) array and a very safe shower muon identification as with CRT detectors. If the detector noise level is too large to reject hadronic showers by a single detected muon, much larger muon detection areas are needed. At energies above about 10–20 TeV hadron rejection factors comparable to those of imaging Cherenkov telescopes can be achieved if punch-through electrons can be well suppressed.

Although the purpose of such arrays is not the detection of a known source like the Crab nebula, a detection of the Crab could help to validate the array performance and seems feasible. The main purpose for low threshold arrays would be the search for variable, episodic, or unexpected sources. Spectra of known sources could also be examined at energies where Cherenkov telescopes inevitably run out of statistics. Particle arrays can be valuable all-sky and whole-year monitoring instruments while imaging Cherenkov telescopes, due to the superior hadron rejection, at least below 10 TeV, remain the instruments of choice for looking for gamma rays from a few gamma-ray source candidates. After having established that TeV gamma-ray sources exist, the need for monitoring instruments like particle detector arrays still arises.

## References

Aglietta, M. *et al.*: 1989, *Nucl. Instr. Meth.* **A277**, 23
Aharonian, F. *et al.*: 1993, in *Proc. 23rd Intern. Cosmic Ray Conf.*, Vol. 4, p. 291
Alexandreas *et al.*: 1991, *Nucl. Instr. Meth.* **A311**, 350
Allen, W. H. *et al.*: 1993, *Phys. Rev. D* **48**, 466
Amenomori, M. *et al.* (Tibet AS$\gamma$ Collaboration): 1993a, *Phys. Rev. D* **47**, 2675
Amenomori, M. *et al.* (Tibet AS$\gamma$ Collaboration): 1993b, in Jones, 1993, p. 219
Baillon, P. *et al.*: 1993, *Astroparticle Physics* **1**, 341
Beaman, J. *et al.*: 1993, *Phys. Rev. D* **48**, 4495
Bernlöhr, K. *et al.*: 1993a, in *Proc. 23rd Intern. Cosmic Ray Conf.*, Vol. 4, p. 199
Bernlöhr, K. *et al.*: 1993b, in *Proc. 23rd Intern. Cosmic Ray Conf.*, Vol. 4, p. 670
Borione, A. *et al.*: 1994, *Nucl. Instr. Meth.* **A346**, 329
Capdevielle, J. N. *et al.*: 1993, in Jones, 1993, p. 545
Catalano, O., Linsley, J., and Scarsi, L.: 1992, *Nucl. Phys. B (Proc. Suppl.)* **28B**, 155
Dingus, B. L. *et al.*: 1988, *Phys. Rev. Lett.* **60**, 1785
Gibbs, K. G. *et al.* (CASA Collaboration): 1988, *Nucl. Instr. Meth.* **A264**, 67
Goret, P. *et al.*: 1993, *Astron. Astroph.* **270**, 401
Heintze, J. *et al.*: 1989, *Nucl. Instr. Meth.* **A277**, 29
Jones, L. (Ed.): 1993, *Very High Energy Cosmic-Ray Interactions, VIIth International Symposium*, Ann Arbor, MI, USA, AIP Conf. Proc. 276
Lewis, D. A. *et al.*: 1993, in *Proc. 23rd Intern. Cosmic Ray Conf.*, Vol. 1, p. 279
Lloyd-Evans, J. *et al.*: 1983, *Nature* **305**, 784
Poirier, J.: 1994, private communication
Poirier, J. *et al.*: 1988, *Nucl. Instr. Meth.* **A264**, 81
Poirier, J. *et al.*: 1993, in Jones, 1993, p. 614
Samorski, M. and Stamm, W.: 1983, *Astroph. J.* **268**, 117
Weekes, T.: 1992, *Space Sci. Rev.* **59**, 315
Yuda, T.: 1994, private communication